\documentclass[aps,prl,twocolumn,superscriptaddress,showpacs]{revtex4}
\usepackage{epsfig}
\usepackage{dcolumn}
\begin{document}

\def\GSI{Gesellschaft f{\"u}r Schwerionenforschung mbH, D-64291 Darmstadt,
Germany}
\def\GANIL{GANIL, CEA et IN2P3-CNRS, F-14076 Caen, France}
\def\IPNO{Institut de Physique Nucl{\'e}aire, IN2P3-CNRS et Universit{\'e}, F-91406
Orsay, France}
\def\LPC{LPC, IN2P3-CNRS, ISMRA et Universit{\'e}, F-14050 Caen, France}
\def\SACLAY{DAPNIA/SPhN, CEA/Saclay, F-91191 Gif sur Yvette, France}
\def\LYON{Institut de Physique Nucl{\'e}aire, IN2P3-CNRS et Universit{\'e}, F-69622
Villeurbanne, France}
\def\NAPOLI{Dipartimento di Scienze Fisiche e Sezione INFN, Univ. 
Federico II,
I-80126 Napoli, Italy}
\def\CATANIA{Dipartimento di Fisica dell' Universit{\`a} and INFN, I-95129
Catania, Italy}
\def\ROSS{Forschungszentrum Rossendorf, D-01314 Dresden, Germany}
\def\WARSAW{A.~So{\l}{}tan Institute for Nuclear Studies, Pl-00681 Warsaw, 
Poland}
\def\CNAM{Conservatoire National des Arts et M{\'e}tiers, 
F-75141 Paris Cedex 03, France}
\def\MOSCOW{Institute for Nuclear Research, 117312 Moscow, Russia}
\def\IFJ{H. Niewodnicza{\'n}ski Institute of Nuclear Physics, Pl-31342 Krak{\'o}w,
Poland}

\title{Isotopic Scaling and the Symmetry Energy\\
in Spectator Fragmentation}

\affiliation{\GSI}
\affiliation{\GANIL}
\affiliation{\LPC}
\affiliation{\IPNO}
\affiliation{\SACLAY}
\affiliation{\LYON}
\affiliation{\CATANIA}
\affiliation{\NAPOLI}
\affiliation{\ROSS}
\affiliation{\WARSAW}
\affiliation{\CNAM}
\affiliation{\IFJ}
\affiliation{\MOSCOW}

\author{A.~Le~F{\`e}vre}            \affiliation{\GSI}
\author{G.~Auger}               \affiliation{\GANIL}
\author{M.L.~Begemann-Blaich}   \affiliation{\GSI}
\author{N.~Bellaize}            \affiliation{\LPC}
\author{R.~Bittiger}            \affiliation{\GSI}
\author{F.~Bocage}              \affiliation{\LPC}
\author{B.~Borderie}            \affiliation{\IPNO}
\author{R.~Bougault}            \affiliation{\LPC}
\author{B.~Bouriquet}           \affiliation{\GANIL}
\author{J.L.~Charvet}           \affiliation{\SACLAY}
\author{A.~Chbihi}              \affiliation{\GANIL}
\author{R.~Dayras}              \affiliation{\SACLAY}
\author{D.~Durand}              \affiliation{\LPC}
\author{J.D.~Frankland}         \affiliation{\GANIL}
\author{E.~Galichet}            \affiliation{\LYON}\affiliation{\CNAM}
\author{D.~Gourio}              \affiliation{\GSI}
\author{D.~Guinet}              \affiliation{\LYON}
\author{S.~Hudan}               \affiliation{\GANIL}
\author{G.~Imm\'{e}}            \affiliation{\CATANIA}
\author{P.~Lautesse}            \affiliation{\LYON}
\author{F.~Lavaud}              \affiliation{\IPNO}
\author{R.~Legrain}\thanks{deceased}            \affiliation{\SACLAY}
\author{O.~Lopez}               \affiliation{\LPC}
\author{J.~{\L}ukasik}             \affiliation{\GSI}\affiliation{\IFJ}
\author{U.~Lynen}               \affiliation{\GSI}
\author{W.F.J.~M{\"u}ller}          \affiliation{\GSI}
\author{L.~Nalpas}              \affiliation{\SACLAY}
\author{H.~Orth}                \affiliation{\GSI}
\author{E.~Plagnol}             \affiliation{\IPNO}
\author{G.~Raciti}              \affiliation{\CATANIA}
\author{E.~Rosato}              \affiliation{\NAPOLI}
\author{A.~Saija}               \affiliation{\CATANIA}
\author{C.~Schwarz}             \affiliation{\GSI}
\author{W.~Seidel}              \affiliation{\ROSS}
\author{C.~Sfienti}             \affiliation{\GSI}
\author{B.~Tamain}              \affiliation{\LPC}
\author{W.~Trautmann}           \affiliation{\GSI}
\author{A.~Trzci\'{n}ski}       \affiliation{\WARSAW}
\author{K.~Turz{\'o}}               \affiliation{\GSI}
\author{E.~Vient}               \affiliation{\LPC}
\author{M.~Vigilante}           \affiliation{\NAPOLI}
\author{C.~Volant}              \affiliation{\SACLAY}
\author{B.~Zwiegli\'{n}ski}     \affiliation{\WARSAW}
\author{A.S.~Botvina}           \affiliation{\GSI}\affiliation{\MOSCOW}
\collaboration{The INDRA and ALADIN Collaborations}
\noaffiliation

\date{\today}

\begin{abstract}

Isotopic effects in the fragmentation of excited target residues following 
collisions of $^{12}$C on $^{112,124}$Sn at incident energies of 300 and 
600~MeV per nucleon were studied with the INDRA 4$\pi$ detector. The 
measured yield ratios for light particles and fragments with atomic number
$Z \leq$ 5 obey the exponential law of isotopic scaling. The deduced scaling
parameters decrease strongly with increasing centrality to values  
smaller than 50\% of those obtained for the peripheral event groups. 
Symmetry term coefficients, deduced from these data within the statistical 
description of isotopic scaling, are near $\gamma =$ 25~MeV for peripheral and 
$\gamma <$ 15~MeV for central collisions.

\end{abstract}


\maketitle

The growing interest in isospin effects in nuclear reactions 
is motivated by an increasing awareness of the importance of the
symmetry term in the nuclear equation of state, in particular for 
astrophysical applications. Supernova simulations
or neutron star models require inputs for the nuclear equation of state at
extreme values of density and asymmetry 
\cite{lattimer,lattprak,botv04}. 
The demonstration in the laboratory of the effects of the symmetry term 
at abnormal densities is, therefore, an essential first step within a 
program aiming at gaining such information experimentally
\cite{bao02,greco02}.

Multifragmentation is generally considered a low-density phenomenon,
with a high degree of thermalization believed to be reached. 
Accepting the concept of a freeze-out volume and the applicability of grand  
canonical logic, the probability of producing a cluster of a given atomic
number $Z$ and mass $A$ at temperature $T$ depends exponentially on the free 
energy of that cluster, $F(Z,A,T)$. The cluster free energies depend 
on the strength of the symmetry term $E_{\rm sym} = \gamma (A-2Z)^2/A$ in the 
liquid-drop energy which, in turn, must depend on the 
extent of expansion of the fragments. This work makes use of an observable 
that isolates the symmetry contribution to the cluster free energy to 
explore the difference in this term for fragments produced in peripheral 
and central collisions. It is found that, 
while the sequential decay strongly degrades the quality
of this observable, the symmetry energy coefficient does indeed 
decrease as the collisions producing the fragments become more violent. 

In the Copenhagen statistical multifragmentation model (SMM), 
standard coefficients $\gamma$ = 23 to 25~MeV are used 
to describe the nascent fragments \cite{bond95,botv02,gamma}. 
In the freeze-out scenario adopted there, normal-density fragments are 
considered to be statistically distributed within an expanded volume,
and the density is only low on average. 
An experimental value for $\gamma$ of about standard magnitude has recently been
obtained within a statistical description of isotopic scaling in light-ion
(p, d, $\alpha$) induced reactions at relativistic energies of up to 15 GeV 
\cite{botv02}. This result, however, may not be representative for 
multi-fragment decays because the data were inclusive and the mean 
multiplicities of intermediate-mass fragments correspondingly small
\cite{beaulieu}. 
In the present work, we apply the same method to exclusive data obtained
with heavier projectiles, $^{12}$C on $^{112}$Sn and $^{124}$Sn targets at 300 
and 600~MeV per nucleon incident energy. Here, according to the 
established systematics \cite{schuett96}, maximum fragment production 
occurs at central impact parameters.

Isotopic scaling, also termed isoscaling, has been shown to be a 
phenomenon common to many different types of heavy ion reactions 
\cite{botv02,tsang01,soul03,fried04}. 
It is observed by comparing product yields 
from otherwise identical 
reactions with isotopically
different projectiles or targets, and it is constituted by an 
exponential dependence of the measured yield ratios $R_{21}(N,Z)$ 
on the neutron number $N$ and proton number $Z$ of the considered product.
The scaling expression
\begin{equation}
R_{21}(N,Z) = Y_2(N,Z)/Y_1(N,Z) = C \cdot exp(\alpha N + \beta Z)
\label{eq:scalab}
\end{equation}
describes rather well the measured ratios over a wide range of
complex particles and light fragments \cite{tsang01a}. 

In the grand-canonical approximation,
assuming that the temperature $T$ is about the same,
the scaling parameters $\alpha$ and $\beta$ are proportional
to the differences of the neutron and proton chemical potentials for
the two systems, $\alpha = \Delta \mu_{\rm n}/T$ and $\beta = \Delta \mu_{\rm p}/T$.
Of particular interest is their connection 
with the symmetry term coefficient. 
It has been obtained from the statistical interpretation of isoscaling 
within the SMM \cite{botv02} and Expanding-Emitting-Source Model 
\cite{tsang01a} and confirmed by an analysis of reaction dynamics
\cite{ono03}. The relation is 
\begin{equation} \label{eq:dmunu}
\alpha T = \Delta \mu_{\rm n} = \mu_{\rm n,2} - \mu_{\rm n,1} \approx 4\gamma
(\frac{Z_{1}^2}{A_{1}^2}-\frac{Z_{2}^2}{A_{2}^2})
\end{equation}
where $Z_{i}$ and $A_{i}$ are the charges and mass
numbers of the two systems 
(the indices 1 and 2 denote the neutron poor and neutron rich system, 
respectively).
With the knowledge of the temperature and the isotopic compositions,
the coefficient $\gamma$ of the symmetry term can be obtained from isoscaling.

The data were obtained with the INDRA multidetector
\cite{pouthas} in experiments performed at the GSI.
Beams of $^{12}$C with 300 and 600~MeV per nucleon incident energy,
delivered by the heavy-ion synchrotron SIS,
were directed onto enriched targets of $^{112}$Sn (98.9\%) and $^{124}$Sn
(99.9\%) with areal densities between 1.0 and 1.2~mg/cm$^2$.
Light charged particles and fragments ($Z \leq 5$) were detected and 
isotopically identified with the calibration telescopes of rings 10 to 17 
of the INDRA detector which cover the range of polar angles 
$45^{\circ} \leq \theta_{\rm lab} \leq 176^{\circ}$. These telescopes consist of pairs of an 80-$\mu$m
Si detector and a 2-mm Si(Li) detector which are mounted
between the ionization chamber and the CsI(Tl) crystal of one of the
modules of a ring \cite{pouthas}. 
Further experimental details may be found in \cite{turzo04} 
and the references given therein. 
For impact-parameter selection, the charged-particle multiplicity 
$M_{\rm C}$ measured with the full 
detector was used, and four bins were chosen for the sorting of the data.
For 600~MeV per nucleon, the two most central bins were combined
for reasons of counting statistics. 

\begin{figure}
     \epsfysize=7.0cm
     \centerline{\epsffile{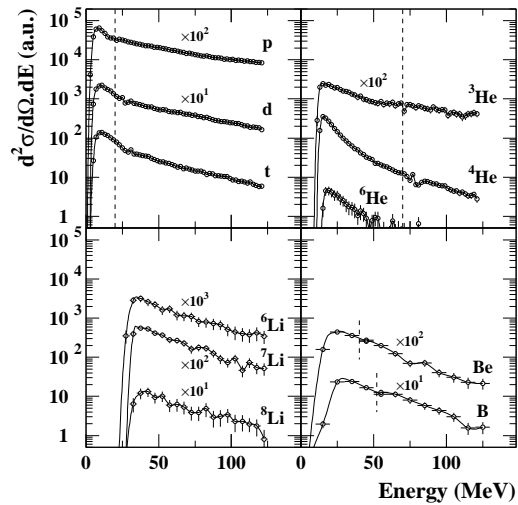}}
\caption{\label{fig:spec}
Energy spectra of H, He, Li isotopes, and of Be and B elements, measured with 
the calibration telescopes for $^{12}$C + $^{124}$Sn at 300~MeV per nucleon. 
The dashed lines indicate the upper limits set for $Z = 1,2$ in the analysis
and the identification thresholds for $Z = 4,5$.
}
\end{figure}

Kinetic energy spectra of light reaction products with $Z \leq 5$, 
integrated over the impact parameter and the angular range $\theta_{\rm lab} \geq 45^{\circ}$,
are shown in Fig.~\ref{fig:spec}.  
To reduce preequilibrium contributions, upper limits of 20~MeV and 70~MeV 
were set for hydrogen and helium isotopes, respectively, which, however, 
are not crucial. The spectra of 
Li, Be, and B fragments were integrated above the energy thresholds 
for isotopic identification which amounted to 28, 40, and 52~MeV,
respectively.

\begin{figure}
     \epsfysize=7.0cm
     \centerline{\epsffile{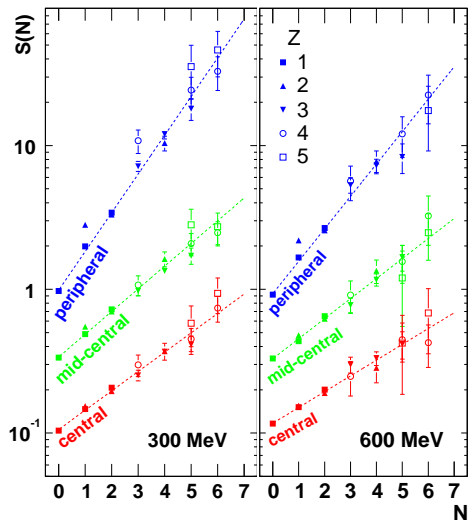}}
\caption{\label{fig:iso}
Scaled isotopic ratios $S(N)$ for $^{12}$C + $^{112,124}$Sn at $E/A$ = 300~MeV
(left panel) and 600~MeV (right panel) for the intervals of reduced 
impact parameter specified in Table~\ref{tab:table1}, with
''central'' indicating $b/b_{\rm max} \leq 0.4$ and with offset factors of 
multiples of three. 
The dashed lines are the results of exponential fits according 
to Eq.~(\protect\ref{eq:scalab}). Only statistical errors are displayed.
}
\end{figure}

\begin{figure}
     \epsfysize=7.0cm
     \centerline{\epsffile{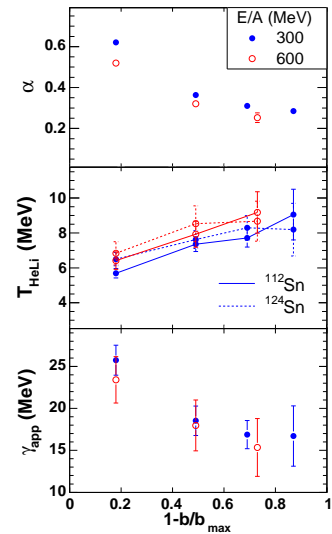}}
\caption{\label{fig:data300600}
Isoscaling coefficient $\alpha$ (top), double-isotope temperatures $T_{\rm HeLi}$ 
(middle) and resulting $\gamma_{\rm app}$ (bottom) for $E/A$ = 300~MeV (full symbols)
and 600~MeV (open symbols), as a function of the centrality 
parameter $1-b/b_{\rm max}$. The temperatures for the $^{112}$Sn and $^{124}$Sn targets 
are distinguished by full and dashed lines, respectively.
}
\end{figure}

The ratios of the fragment yields measured for the 
two reactions and integrated over the chosen intervals of energy and 
angle ( $\theta_{\rm lab} \geq 45^{\circ}$) obey the law of isoscaling. 
This is illustrated in Fig.~\ref{fig:iso} which shows the scaled 
isotopic ratios $S(N) = R_{21}(N,Z)/{\rm exp}(\beta Z)$.
Their slope parameters change considerably with impact parameter, 
extending from $\alpha$~=~0.62 to values as low as 
$\alpha$~=~0.25 for the most central event group at 600~MeV per nucleon
(Table~\ref{tab:table1} and Fig.~\ref{fig:data300600}, top). 

\begin{table}
\caption{\label{tab:table1}
Parameters obtained from fitting the measured isotopic
yield ratios with the scaling function given in
Eq.~(\protect\ref{eq:scalab}).
}
\begin{ruledtabular}
\begin{tabular}{l c c c c }
 & 300 MeV & & 600 MeV & \\
 $b/b_{\rm max}$ & $\alpha$ & $\beta$ & $\alpha$ & $\beta$ \\
\hline
 0.0 - 0.2 &
 0.28 $\pm$ 0.01 &
-0.33 $\pm$ 0.03 &
               &
                \\
 0.2 - 0.4 &
 0.31 $\pm$ 0.01 &
-0.32 $\pm$ 0.01 &
 0.25 $\pm$ 0.02 &
-0.28 $\pm$ 0.04 \\
 0.4 - 0.6 &
 0.36 $\pm$ 0.01 &
-0.39 $\pm$ 0.02 &
 0.32 $\pm$ 0.02 &
-0.34 $\pm$ 0.03 \\
 0.6 - 1.0 &
 0.62 $\pm$ 0.01 &
-0.68 $\pm$ 0.02 &
 0.52 $\pm$ 0.02 &
-0.59 $\pm$ 0.03 \\
\end{tabular}
\end{ruledtabular}
\end{table}

Temperature estimates were obtained from the yields of $^{3,4}$He and $^{6,7}$Li
isotopes, and the deduced $T_{\rm HeLi}$ contains a correction factor 1.2 for the 
effects of sequential decay \cite{poch95,xi97}. The temperatures 
are quite similar for the two target cases and 
increase with centrality from about 6~MeV to 9~MeV 
(Fig \ref{fig:data300600}, middle).
This is consistent with the results obtained for $^{197}$Au fragmentations 
\cite{poch95,xi97} and with the established dependence on the system 
mass \cite{nato02}. The rise of $T_{\rm HeLi}$, however,
does not compensate for the decrease of $\alpha$, as it did in the case
of light-particle induced reactions \cite{botv02}, and $\Delta \mu_{\rm n}$, consequently,
decreases toward the central collisions.

The analytical expression for $\Delta \mu_{\rm n}$ (Eq.~\ref{eq:dmunu}) contains the
isotopic compositions of the sources, more precisely the difference 
of the squared $Z/A$ values, $\Delta (Z^2/A^2) = (Z_{1}/A_{1})^2 - (Z_{2}/A_{2})^2$.
For the target spectators, this quantity is not expected to deviate 
significantly from its original value \cite{botv02}, 
in contrast to mean-field dominated reaction systems
at intermediate energies \cite{ono03,shetty04}.
This was confirmed with calculations performed with the 
Li{\`e}ge-cascade-percolation model \cite{volant04} 
and with the Relativistic Mean Field Model (RBUU, Ref. \cite{gait04})
in which isotopic effects of the nuclear mean field are treated explicitly.
The individual $Z/A$ values change slightly but $\Delta (Z^2/A^2)$ remains 
nearly the same. For central collisions at 600~MeV per nucleon, e.g., 
the RBUU calculations predict a reduction of $\Delta (Z^2/A^2)$ by 6\% for the 
target-rapidity region after 90 fm/c collision time. 

The suggested corrections are small and 
can thus be temporarily ignored.
With the compositions of the original targets, $\Delta (Z^2/A^2)$ = 0.0367, the
expression $\gamma = \alpha T/0.147$ is obtained from Eq.~\ref{eq:dmunu}. 
It was used to determine an apparent 
symmetry-term coefficient
$\gamma_{\rm app}$, i.e. without sequential decay corrections for $\alpha$, 
from the data shown in Fig.~\ref{fig:data300600} (the 
mean values were used for $T$).
The results are close to the normal-density coefficient for peripheral 
collisions but drop to lower values at the more central impact parameters 
(Fig.~\ref{fig:data300600}, bottom).

The effects of sequential decay were studied with
the microcanonical Markov-chain version of the Statistical Multifragmentation
Model \cite{botv01}. The target nuclei $^{112,124}$Sn with excitation energies 
of 4, 6, and 8~MeV per nucleon were chosen as inputs, 
and the symmetry term $\gamma$ was varied between 4 and 25~MeV. 
The isoscaling coefficient $\alpha$ was 
determined from the calculated fragment yields before (hot fragments) and 
after (cold fragments) the sequential decay stage of the calculations
for which standard values for the fragment masses were used.
The energy balance at freeze-out and during the secondary 
deexcitation was taken into account as described in \cite{bond95}.

\begin{figure}
     \epsfysize=7.0cm

     \centerline{\epsffile{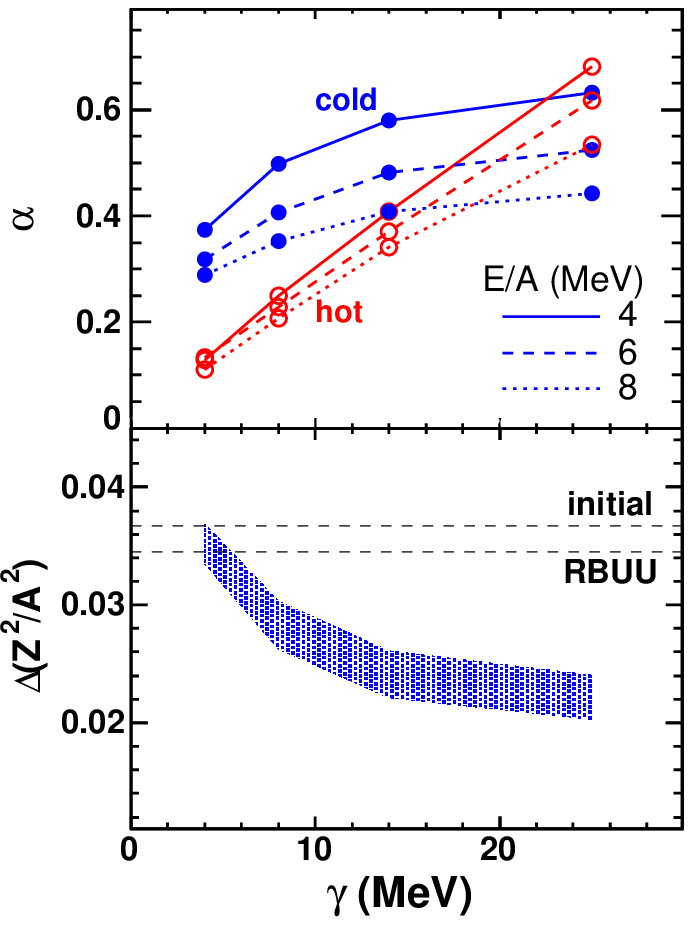}}

\caption{\label{fig:symm}
Top panel: isoscaling coefficient $\alpha$ for hot (open circles) and cold 
fragments (dots) as a function of the symmetry term coefficient $\gamma$ 
as predicted by the Markov-chain calculations for $^{112,124}$Sn.\\
Bottom panel: region in the $\Delta (Z^2/A^2)$-versus-$\gamma$ plane
consistent with the measured value $\alpha =0.29$ for central collisions
and with the Markov-chain predictions for cold fragments (shaded).
The dashed lines indicate the $\Delta (Z^2/A^2) = 0.0367$ of $^{112,124}$Sn and the 
RBUU prediction.
}
\end{figure}

The hot fragments exhibit the linear relation of $\alpha$ with $\gamma$ as expected
(Fig.~\ref{fig:symm}, top panel).
With $\gamma$ = 25~MeV, the sequential processes cause a slight broadening 
of the isotopic distributions and the resulting $\alpha$ is lowered by 10\% to 20\%, 
similar to what was reported in \cite{botv02}. 
For smaller values of $\gamma$, however, larger changes of $\alpha$ are predicted.
The decay of the wings of the wider distributions of hot fragments 
toward the valley of stability causes the resulting cold distributions 
to be narrower and the isoscaling coefficients to be larger.
The overall variation of $\alpha$ with $\gamma$ is weaker, and 
the decrease of $\gamma$ with centrality should, thus, be stronger 
than that of $\gamma_{\rm app}$ (Fig.~\ref{fig:data300600}). For central collisions,
this implies $\gamma <$ 15~MeV as an upper limit but much smaller values 
are more likely to be expected. With the present calculations, 
for excitation energies up to 8~MeV per nucleon and the original 
target compositions, the measured $\alpha < 0.3$  
is only reproduced with $\gamma \approx$ 6 MeV (Fig.~\ref{fig:symm}, top).

The bottom panel of Fig.~\ref{fig:symm} shows how the situation changes when 
the variation of the isotopic compositions of the two systems is again 
included as a degree of freedom. 
The shaded band in the plane of $\Delta (Z^2/A^2)$ versus $\gamma$ 
represents the region consistent with the 
weighted mean $\alpha = 0.29$ measured for the central bins ($b/b_{\rm max} \leq 0.4$) 
at the two energies.
It was obtained from the predictions $\alpha(\gamma)$ for cold fragments at 
excitation energies between 6 and 8~MeV per nucleon 
(Fig.~\ref{fig:symm}, top)
according to $\Delta (Z^2/A^2) = 0.0367 \cdot 0.29/ \alpha(\gamma)$, 
i.e. using that $\alpha$ is, to first order, proportional to the difference
of the isotopic compositions, here expressed 
by $\Delta (Z^2/A^2)$ (see, e.g., \cite{botv02,ono03}).
If this difference remains close to the
cascade and RBUU predictions the resulting 
symmetry term coefficient for the central reaction channels is very small, 
$\gamma \leq$ 10~MeV. 
To restore the compatibility with a standard $\gamma \approx$ 25~MeV 
would require considerable isotopic asymmetries 
in the initial reaction phase, much larger than what is   
expected according to the models.

In conclusion, the observed decrease of the isoscaling parameter $\alpha$ with
centrality 
which is not compensated by a correspondingly increasing temperature 
requires a decreasing symmetry term coefficient $\gamma$ in a statistical 
description of the fragmentation process. The effect is enhanced 
if sequential fragment decay is taken into account. 
Values less than 15~MeV, 
as obtained from the present analysis, are not necessarily
unreasonable in a realistic description of the chemical freeze-out state.
Besides the global expansion of the system also the possibly 
expanded or deformed structure of the forming fragments as well as 
their interaction with other fragments and with the surrounding nucleon gas 
will have to be considered. The presented results depend, crucially, 
on the isotopic evolution of the multi-fragmenting system as it approaches 
the chemical freeze-out and, more quantitatively, on the treatment of 
sequential decay in the analysis. These questions deserve further attention.

The authors would like to thank T. Gaitanos for communicating the results of
RBUU calculations and for valuable discussions.
This work was supported by the European Community under
contract ERBFMGECT950083.


\begin{thebibliography}{99}
\bibitem{gamma}
the notation follows that originally chosen in Ref. \cite{bond95}; 
alternatively $C_{\rm sym}$ is frequently used for the same quantity.
\end{thebibliography}
\end{document}